\documentclass{ieeeaccess}
\usepackage{cite}
\usepackage{amsmath,amssymb,amsfonts}
\usepackage{algorithm,algorithmic}
\usepackage{graphicx}
\usepackage{textcomp}

\def\BibTeX{{\rm B\kern-.05em{\sc i\kern-.025em b}\kern-.08em
    T\kern-.1667em\lower.7ex\hbox{E}\kern-.125emX}}
\begin{document}
\history{Received October 27, 2020, accepted November 18, 2020, date of publication November 24, 2020,\\ date of current version December 7, 2020.}
\doi{10.1109/ACCESS.2020.3040017}

\title{Guiding Principle for Minor-Embedding in Simulated-Annealing-Based Ising Machines}
\author{\uppercase{Tatsuhiko Shirai}\authorrefmark{1}, 
\uppercase{Shu Tanaka}\authorrefmark{2,3}, and \uppercase{Nozomu Togawa},\authorrefmark{1}
\IEEEmembership{Member, IEEE}}
\address[1]{Department of Computer Science and Communications Engineering, Waseda University, Tokyo 169-8555, Japan}
\address[2]{Department of Applied Physics and Physico-Informatics, Keio University, Kanagawa 223-8522, Japan}
\address[3]{Green Computing System Research Organization, Waseda University, Tokyo 162-0042, Japan}
\tfootnote{
}

\markboth
{Tatsuhiko Shirai \headeretal: Guiding Principle for ME in Simulated-Annealing-Based Ising Machines}
{Tatsuhiko Shirai \headeretal: Guiding Principle for ME in Simulated-Annealing-Based Ising Machines}

\corresp{Corresponding author: Tatsuhiko Shirai (e-mail: tatsuhiko.shirai@aoni.waseda.jp).}

\begin{abstract}
We propose a novel type of minor-embedding (ME) in simulated-annealing-based Ising machines.
The Ising machines can solve combinatorial optimization problems.
Many combinatorial optimization problems are mapped to find the ground (lowest-energy) state of the logical Ising model.
When connectivity is restricted on Ising machines, ME is required for mapping from the logical Ising model to a physical Ising model, which corresponds to a specific Ising machine.
Herein we discuss the guiding principle of ME design to achieve a high performance in Ising machines.
We derive the proposed ME based on a theoretical argument of statistical mechanics.
The performance of the proposed ME is compared with two existing types of MEs for different benchmarking problems.
Simulated annealing shows that the proposed ME outperforms existing MEs for all benchmarking problems, especially when the distribution of the degree in a logical Ising model has a large standard deviation.
This study validates the guiding principle of using statistical mechanics for ME to realize fast and high-precision solvers for combinatorial optimization problems.
\end{abstract}

\begin{keywords}
Annealing machine, graph minor-embedding, Ising model, optimization method, simulated annealing, statistical mechanics
%Enter key words or phrases in alphabetical 
%order, separated by commas. For a list of suggested keywords, send a blank 
%e-mail to keywords@ieee.org or visit \underline
%{http://www.ieee.org/organizations/pubs/ani\_prod/keywrd98.txt}
\end{keywords}

\titlepgskip=-15pt

\maketitle

\section{Introduction}
\label{sec:introduction}
\subsection{Motivation}
Combinatorial optimization problems find the optimal combination of decision variables to minimize or maximize the objective function under given constraints.
Solving a combinatorial optimization problem with a large number of decision variables is difficult because the number of solution candidates increases exponentially with the number of decision variables.
Typical examples of combinatorial optimization problems found in textbooks include the satisfiability problem, the traveling salesman problem, and the knapsack problem.
In our daily life, combinatorial optimization problems are ubiquitous.
Common examples include the shift-planning optimization, the logistics optimization, and the traffic route optimization.
Consequently, the development of efficient solvers for combinatorial optimization problems has attracted attention both in academia and in industry.

Ising machines have been developed as fast and high-precision solvers for combinatorial optimization problems~\cite{johnson2011quantum, bunyk2014architectural, yamaoka2016a20kspin, inagaki2016coherent, mcmahon2016fully, maezawa2017design, okuyama2017ising, yoshimura2017implementation, aramon2019physics, goto2019combinatorial, maezawa2019toward,okuyama2019binary}.
They employ three phases to solve problems.
In the first phase, a combinatorial optimization problem is mapped as an Ising problem.
The Ising problem finds the ground (lowest-energy) state of the logical Ising model, which was originally introduced in statistical mechanics to describe the nature of phase transition materials~\cite{nishimori2010elements,nakahara2013lectures}.
The Ising model consists of spins with values of $+1$ or $-1$.
As described in Sec.~\ref{Subsec:Embedding}, the logical Ising model is defined on an undirected graph with unrestricted connectivity between vertices.
The objective function and the constraints in a combinatorial optimization problem are encoded in the Ising model~\cite{lucas2014Ising, tanaka2017quantum, tanahashi2019application}.
Different encoding methods have been proposed: machine learning~\cite{neven2008training}, portfolio optimization~\cite{rosenberg2016solving,tanahashi2019application}, traffic optimization~\cite{neukart2017traffic}, optimization in an integrated design circuit~\cite{kanamaru2019efficient, terada2018ising}, and material design~\cite{kitai2019expanding}.
In the second phase, the logical Ising model formulated in the first phase is mapped onto a physical Ising model. The model corresponds to the Ising machine considered.
Here, the physical Ising model is defined on an undirected graph where the connectivity between vertices may be restricted.
For Ising machines with restricted connectivity such as D-Wave~\cite{johnson2011quantum, bunyk2014architectural} and CMOS annealing machines~\cite{yamaoka2016a20kspin, yoshimura2017implementation}, the mapping called minor-embedding (ME)~\cite{choi2008minor} is necessary.
In ME, a single spin in the logical Ising model is expressed by several spins in the physical Ising model.
The set of spins is called a chain since chains are often formed in an actual ME.
In the third phase, the Ising machine searches for the lowest-energy state according to its operation principle.

ME can be classified into two types according to the number of spins in each chain.
In the first type, each chain has the same number of spins (i.e., a uniform chain length).
This type of ME is often called clique ME or complete-graph ME because the logical Ising model with all-to-all coupling can be embedded.
The algorithms for this type of ME have been developed for D-Wave~\cite{choi2011minor,klymko2014adiabatic,boothby2016fast} and CMOS annealing machines~\cite{oku2019fully}.
In the second type, each chain has a different number of spins.
The total number of spins in the physical Ising model is usually smaller in the second type if the logical Ising model is not fully connected.
Thus, the second type can embed a larger number of logical spins.
Heuristic algorithms for finding this type of ME have been developed~\cite{cai2014practical,perdomo2015performance,hamilton2017identifying,zaribafiyan2017systematic,harris2018phase,king2018observation,sugie2018graph,okada2019improving}.
In existing MEs, the spins in a chain interact with ferromagnetic coupling.
In both types of MEs, the chains have the same coupling strength.
Here, we call the two types of MEs ``uniform-length and uniform-coupling ME (ME~$i$)'' and ``nonuniform-length and uniform-coupling ME (ME~$ii$)'', respectively.

\subsection{Summary of Contributions}
Herein, we discuss the guiding principle of ME design to achieve a higher performance in simulated-annealing (SA)-based Ising machines.
The main contributions are:

\begin{itemize}
    \item 
    A novel type of ME is proposed where the lengths are nonuniform and the coupling strength of each chain depends on the chain length.
    The formula between the coupling strength and the chain length is derived from a viewpoint of statistical mechanics.
    The coupling strength increases with the chain length.
    This type of ME, which is herein called ``nonuniform-length and nonuniform-coupling ME (ME~$iii$)'', has not been discussed in previous studies.\\
    
    \item 
    The performance of our proposed ME is compared to two existing types of ME through SA.
    The results demonstrate that the proposed ME has the best performance for all the problems.
    In particular, it outperforms the others when the degree of the logical Ising model is widely distributed.
    The results are general and independent of the distribution of the coupling strengths and biases in logical Ising models.\\
    \end{itemize}

The rest of the paper is organized as follows.
Section~\ref{Sec:SA} briefly introduces the SA and thermal equilibrium states.
The idea of thermal equilibrium states is necessary to derive the proposed ME (ME~$iii$).
Section~\ref{Sec:Embedding} discusses ME to fix the notation.
Then a physical Ising model is presented to tune the chain lengths and intra-chain-coupling strengths in ME.
With this model, we show the new type of ME as well as the two existing types of ME.
Section~\ref{Sec:Experiment} explains the experimental setup.
Section~\ref{Numerical} demonstrates the numerical results, and Sec.~\ref{Sec:Discussion} discusses our results.
Section~\ref{Sec:conclusion} concludes with a summary of the results and future research directions.
The Appendices give supplemental information for the derivation of the proposed ME (Appendix~\ref{Appendix1}) and the experimental results (Appendix~\ref{Appendix2} and Appendix~\ref{Appendix3}).

\section{Simulated Annealing and the thermal equilibrium state}\label{Sec:SA}
SA is a heuristic algorithm.
It is useful in a wide range of application~\cite{kirkpatrick1983optimization, johnson1989optimization,johnson1991optimization}.
It has been employed to find the optimal solution of an objective function in combinatorial optimization problems.
To explain SA as an operation principle of Ising machines in the language of statistical mechanics, we consider the objective function as an energy function, which is referred to as the Hamiltonian of the Ising model.
As explained in Sec.~\ref{sec:introduction}, the Ising model consists of spins with values of $+1$ and $-1$.
Let $H(\{\sigma_i\})$ be the Hamiltonian of the Ising model, where $\{\sigma_i\}$ is a combination of decision variables called spins.
In this case, the ground (lowest-energy) state corresponds to the spin combination (spin configuration) $\{\sigma_i\}$ that minimizes the value of $H(\{\sigma_i\})$.

\begin{algorithm}[t]
 \caption{Simulated annealing implemented by Markov Chain Monte Carlo}
 \begin{algorithmic}[1]
  \FOR {each run}
  \STATE initialize to a random initial state
  \FOR {each temperature $T$}
  \FOR {each Monte Carlo sweep at the temperature}
  \STATE choose a candidate site
  \STATE calculate the energy difference $\Delta E$ given by (\ref{energy_diff})
  \STATE generate a random number $r$ such that $0\leq r < 1$
  \STATE if $r$ is less than the transition probability $W(\Delta E, T)$, update the state
  \ENDFOR
  \STATE lower the temperature
  \ENDFOR
  \ENDFOR
 \end{algorithmic} 
 \label{SA}
 \end{algorithm}

Algorithm~\ref{SA} shows the SA algorithm implemented by Markov Chain Monte Carlo (MCMC).
The algorithm starts from a completely random initial state.
That is, the spin configuration $\{\sigma_i\}$ is arbitrarily selected.
Then the spin configuration is repeatedly updated.
Let us consider a transition from the current state $\{ \sigma'_i\}$ to a candidate state $\{ \sigma_i \}$.
The probability of making the transition is specified by a transition probability $W(\Delta E, T)$, which depends on temperature $T$ and the energy difference between the two states defined by
\begin{equation}
\Delta E= H(\{ \sigma_i \}) - H( \{ \sigma'_i \}).
\label{energy_diff}
\end{equation}
According to the principle of MCMC, the transition probability $W(\Delta E, T)$ must satisfy the balance condition, which is given as
\begin{equation}
\sum_{\{\sigma_i\}} W(\Delta E, T) {\rm P}^\rm{eq}_T (\{\sigma'_i \}) = \sum_{\{\sigma_i\}} W(-\Delta E, T) {\rm P}^\rm{eq}_T (\{\sigma_i \}),
\label{detailed}
\end{equation}
where the summation means the summation of all the spin configurations.
Two well-known choices of the transition probability satisfying the above equation are the heat-bath method and the Metropolis method.
Here, ${\rm P}_T^\rm{eq} (\{\sigma_i \})$ is the probability distribution of the thermal equilibrium state at temperature $T$ and is given by
\begin{equation}
{\rm P}^\rm{eq}_T (\{\sigma_i \}) = \frac{\exp{\left[-\frac{H(\{\sigma_i\})}{T}\right] }}{\sum_{\{\sigma'_i\}} \exp \left[ -\frac{H(\{\sigma'_i\})}{T} \right] }.
\end{equation}
Here, we set the Boltzmann constant, which is a physical constant, to unity.
When the temperature $T$ is fixed,  Algorithm~\ref{SA} is used to sample spin configurations in a thermal equilibrium state~\cite{landau2014guide,tanaka2017quantum}.
The thermal equilibrium state at high temperature is a random state where the population is almost the same for all spin configurations.
By contrast, the thermal equilibrium state at low temperature has a large population in the lower-energy states.
In SA, by gradually lowering the temperature, the state should make transitions from a high-temperature state to a low-temperature state while annealing.
After performing SA, a lower-energy state, ideally the ground state of $H(\{ \sigma_i \})$, is identified.

Herein the expectation value of a physical quantity in the thermal equilibrium state at temperature $T$ is referred to as the thermal average and is denoted as
\begin{equation}
\langle f \rangle_T :=\sum_{\{\sigma_i\}} f(\{\sigma_i\}) {\rm P}_T^\rm{eq} (\{\sigma_i\}; T),
\end{equation}
where $f(\{\sigma_i\})$ is an arbitrary function of the spin configuration.

\section{Minor-Embedding}\label{Sec:Embedding}
In this section, we describe our proposed ME.
First, we introduce the concept of ME and a physical Ising model to systematically tune the chain length and the intra-chain-coupling strength.
%propose a new type of ME ``un-uniform-length and un-uniform-coupling ME (ME~$iii$)'' in Sec.~\ref{Subsec:algorithm}.
%Before the proposal of the ME, we start with a brief introduction of ME in Sec.~\ref{Subsec:Embedding} and introduce a physical Ising model for systematically tuning the chain length and the intra-chain-coupling strength in Sec.~\ref{Subsec:model}.
%Sec.~\ref{Appendix0} gives supplemental information for the derivation of the proposed embedding.

\subsection{Brief introduction of Minor-Embedding}\label{Subsec:Embedding}
ME is the mapping from a logical Ising model to a physical Ising model.
The symbols L and P denote a logical Ising model and a physical Ising model, respectively. 
The logical Ising model is defined on an undirected graph $G_{\rm L}=(V_{\rm L}, E_{\rm L})$, where $V_{\rm L}$ and $E_{\rm L}$ are sets of vertices and edges, respectively.
Herein we refer to $G_{\rm L}$ as a logical graph.
The number of vertices is denoted by $N_{\rm L}$.
As mentioned in Sec.~\ref{Sec:SA}, the Hamiltonian of the Ising model is an objective function and is given by
\begin{equation}
H_{\rm L}(\{ \sigma_i \})=-\sum_{(i,j)\in E_{\rm L}} J_{ij} \sigma_i \sigma_j -\sum_{i\in V_{\rm L}} h_i \sigma_i,
\end{equation}
where $\sigma_i \in \{-1,1\}$ is the logical spin, $J_{ij}$ is the interaction between spins $\sigma_i$ and $\sigma_j$, and $h_i$ is the bias on the spin $\sigma_i$.
Both $J_{ij}$ and $h_i$ are real values.
$J_{ij} >0$ indicates ferromagnetic coupling, whereas $J_{ij}<0$ denotes antiferromagnetic coupling.
Many combinatorial optimization problems can be mapped as problems to find the ground state of $H_{\rm L}(\{ \sigma_i \})$.
The interaction strengths $J_{ij}$ and the biases $h_i$ are specified by the objective function and constraints of the combinatorial optimization problem.

In a similar manner, the physical Ising model is defined on an undirected graph $G_{\rm P}=(V_{\rm P}, E_{\rm P})$, where a physical spin with a binary variable is put on each vertex.
[For the specific form of the Hamiltonian in this study, see eq.~(\ref{physical_Ising}).]
Hereafter, $G_{\rm P}$ is referred to as the physical graph and it corresponds to the graph determined by the Ising machine architecture.
In general, the physical graph $G_{\rm P}$ has a degree constraint where each vertex can have at most a constant degree.
For example, the degree is $6$ for the Chimera graph~\cite{bunyk2014architectural}, $15$ for the Pegasus graph in the D-Wave machines~\cite{dattani2019pegasus}, and $5$ (1st generation prototype~\cite{yamaoka2016a20kspin}) and $8$ (2nd generation prototype~\cite{yoshimura2017implementation}) in the CMOS annealing machines.
Due to the connectivity restriction among vertices, the logical graph $G_{\rm L}$ is not typically a subgraph of $G_{\rm P}$.

ME enables $G_{\rm L}$ to be expressed in $G_{\rm P}$ even when $G_{\rm L}$ is not a subgraph of $G_{\rm P}$.
Each vertex in the logical graph, $i \in V_{\rm L}$, is mapped to a set of several vertices in the physical graph, $\phi(i) \subset V_{\rm P}$.
ME is defined by mapping $\phi: V_{\rm L} \to V_{\rm P}$, which satisfies the following conditions~\cite{cai2014practical}:
\begin{enumerate}
\item For each vertex $i \in V_{\rm L}$, the vertices in $\phi(i) \subset V_{\rm P}$ are connected and the connection is called chain;
\item For all $i\neq j$ in $V_{\rm L}$, $\phi(i)$ and $\phi(j)$ are disjointed;
\item For each pair $(i, j) \in E_{\rm L}$, the corresponding pair exists in the physical graph (i.e., a pair of vertices, $k \in \phi(i)$ and $\ell \in \phi(j)$, satisfying $(k, \ell)\in E_\rm{P}$).
\end{enumerate}
The physical spins in a chain interact with a ferromagnetic coupling.
When the ferromagnetic coupling is sufficiently large, the ground state of the logical Ising model and that of the physical Ising model have a one-to-one correspondence~\cite{choi2008minor}.
This implies that the ground state of $H_{\rm L}(\{\sigma_i\})$ is obtained by searching the ground state of the embedded physical Ising model.

\subsection{Physical Ising model to tune chain lengths and intra-chain-coupling strengths in Minor-Embedding}\label{Subsec:model}

\begin{figure}[!t]
\centering
\includegraphics[width=\linewidth]{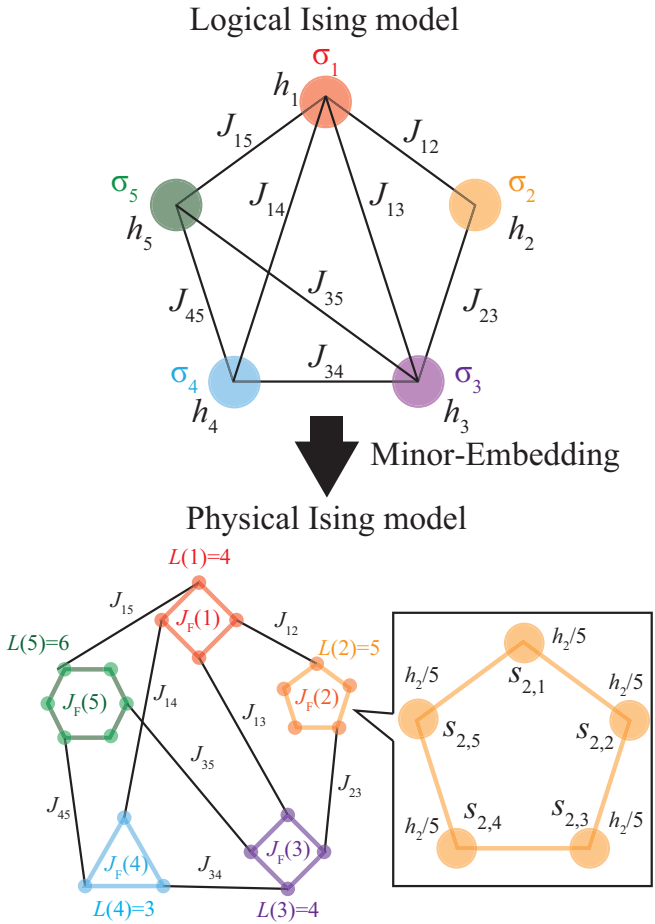}
\caption{
Example of mapping from a logical Ising model to a physical Ising model.
(Upper) Logical Ising model with logical spins $\{\sigma_i \}_{i=1}^5$.
$\{J_{ij}\}$ and $\{ h_i\}$ denote the coupling strengths and the biases in the logical Ising model, respectively.
(Lower) Physical Ising model, which is mapped by the ME of the logical Ising model.
Each spin in the logical Ising model $\sigma_i$ is mapped to a ring with length $L(i)$ and ferromagnetic coupling strength $J_{\rm F}(i)$.
Physical spins are labeled by $s_{i,k}$, where $i$ is the ring label and $k$ runs from $1$ to $L(i)$.
This model can tune the length of the ring $L(i)$ and the intra-ring-coupling strength $J_{\rm F}(i)$.
}
\label{fig_embed}
\end{figure}

This subsection describes a physical Ising model to systematically tune the chain lengths and the intra-chain-coupling strengths.
The upper panel of Fig.~\ref{fig_embed} represents the logical Ising model with $N_{\rm L}=5$.
The logical spins $\sigma_i$ and $\sigma_j$ are connected with coupling strength $J_{ij}$ when there is an edge between the corresponding vertices, and the bias with the strength $h_i$ is applied on each spin $i$.
The lower panel shows the physical Ising model in which the logical Ising model is embedded.
For simplicity, we assume that each chain is a ring of vertices in $G_{\rm P}$.
Each vertex in the logical graph, $i \in V_{\rm L}$, is mapped to the ring with the length $L(i)$, and the physical spins in the ring are connected through a ferromagnetic coupling with the strength $J_{\rm F}(i)$. 
The Hamiltonian of the physical Ising model is explicitly given by
\begin{align}
H_{\rm P} (\{s_{i,j}\})=&-\sum_{(i,j)\in E_{\rm L}} J_{ij} s_{i,v_i(j)} s_{j,v_j(i)} -\sum_{i \in V_{\rm L}} \frac{h_i}{L(i)} \sum_{k=1}^{L(i)} s_{i,k} \nonumber\\
&-\sum_{i \in V_{\rm L}} J_{\rm F} (i) \sum_{k=1}^{L(i)} s_{i,k} s_{i,k+1}, \label{physical_Ising}
\end{align}
where $s_{i, j} \in \{-1, 1\}$ is the $j$-th physical spin in the ring $\phi (i)$ and the periodic boundary condition is imposed (i.e., $s_{i, L(i)+1}=s_{i,1}$).
There is an interaction with the strength $J_{ij}$ between a physical spin in a ring $\phi(i)$ and a physical spin in a ring $\phi(j)$, and $v_i(j)$ denotes the physical spin in a ring $\phi(i)$.
We assume that each physical spin interacts with one physical spin in other rings, at most.
Therefore, $v_i(j) \neq v_i(k)$ if $j \neq k$.
The bias on each physical spin in a ring $\phi(i)$ is set as $h_i/L(i)$.
In this way, the biases applied to the spins in a ring become uniform.
Since intra-ring-couplings are ferromagnetic couplings, $J_{\rm F}(i) > 0$ for all $i$.
When the strength of $J_{\rm F}(i)$ is sufficiently large, the ground states of the logical Ising model and the physical Ising model have a one-to-one correspondence.

\subsection{Types of Minor-Embedding}\label{Subsec:algorithm}
We consider three types of ME: ME~$i$, ME~$ii$, and ME~$iii$.
These depend on the choice of the ring length $L(i)$ and the intra-ring-coupling strength $J_{\rm F}(i)$.
ME~$i$ and ME~$ii$ have been studied previously~\cite{choi2011minor,klymko2014adiabatic,boothby2016fast,oku2019fully,cai2014practical,hamilton2017identifying,zaribafiyan2017systematic,harris2018phase,king2018observation,sugie2018graph,okada2019improving}.
ME~$iii$ is a new type proposed in this study.
Equation~(\ref{physical_Ising}) can systematically express the three types of ME.\\
%The last embedding algorithm is derived based on a theoretical argument in statistical mechanics, which is for the first time obtained in the present paper.

\begin{itemize}
\item{\it ME~i: uniform-length and uniform-coupling ME}\\
In the first type of ME, all rings have the same length and coupling strength.
We set the number of spins in a ring as $N_{\rm L}-1$, which is the maximum degree for each vertex.
That is
\begin{equation}
L(i)=N_{\rm L}-1.    
\end{equation}
We set $v_i(j)$ as
\begin{equation}
v_i(j)=\left\{
\begin{aligned}
&j \qquad\text{if } i > j,\\
&j-1 \text{ if } i < j.
\end{aligned}
\right.
\end{equation}
In ME~$i$, the logical Ising model with all-to-all coupling can be embedded.
We introduce a hyperparameter $J_{\rm c}$ for the intra-ring-coupling, which is expressed as
\begin{equation}
J_{\rm F}(i)=J_{\rm c}.
\end{equation}
%In this embedding algorithm, the graph $G_{\rm P}$ is independent of the graph $G_{\rm L}(E_{\rm L}, V_{\rm L})$.
The number of vertices in the physical graph (i.e., the number of spins in the physical Ising model) is provided as 
\begin{equation}
    |V_{\rm P}|=N_{\rm L}(N_{\rm L}-1).
\end{equation}\\

\item{\it ME~ii: nonuniform-length and uniform-coupling ME}\\
In the second type of ME, the total number of physical spins is set as small as possible.
For a given logical Ising model, it is sufficient to take the number of spins in a ring $\phi(i)$ as the degree of vertex denoted by $k_i$.
Namely,
\begin{equation}
L(i)=k_i.
\end{equation}
We set $v_i (j)$ as
\begin{equation}
v_i(j)=n_i(j) \text{ for } (i,j)\in E_{\rm L},
\label{heuristic}
\end{equation}
where $n_i(j) \in \mathbb{N}$ is an integer given for vertex $i \in V_{\rm L}$.
The integer is incremented by one when there is an edge between the logical spins $i$ and $j$.
That is, $(i,j)\in E_{\rm L}$.
For example, if a logical spin labeled by $2$ interacts with spins labeled by $1$, $5$, and $6$, then $n_2(1)=1$, $n_2(5)=2$, and $n_2(6)=3$.
Similar to ME~$i$, a uniform ferromagnetic coupling strength is assumed inside the ring and
\begin{equation}
J_{\rm F}(i)=J_{\rm c}.
\end{equation}
In this ME, every physical spin is connected to a spin in another ring.
Hence, the number of vertices in the physical Ising model is given by
\begin{equation}
    |V_{\rm P}|=\sum_{i\in V_{\rm L}} \frac{k_i}{2}.
    \label{vertex_Power}
\end{equation}\\

\item{\it ME~iii: nonuniform-length and nonuniform-coupling ME}\\
We propose a new type of ME where the intra-ring-coupling strength depends on the ring length.
Similar to ME~$ii$, the length of ring $L(i)$ is equal to the degree of the vertices $i \in V_{\rm L}$,
\begin{equation}
L(i)=k_i,
\end{equation}
and $v_i (j)$ is set by eq.~(\ref{heuristic}).
The intra-ring-coupling strength $J_{\rm F}(i)$ is given by
\begin{equation}
J_{\rm F}(i)= -\frac{J_{\rm c}}{2} \log \left[ \tanh \left(\frac{1}{2L(i)} \right) \right].
\label{relation}
\end{equation}
%where $T_{\rm c}$ is a hyper-parameter that controls the coupling strength.
Here, $J_{\rm F}(i)$ is a monotonically increasing function of $L(i)$, and it asymptotically behaves as
\begin{equation}
J_{\rm F}(i) \sim \log L(i) \text{ for } L(i) \gg 1.
\label{Log}
\end{equation}
%Therefore, we set larger values of $J_{\rm F}(i)$ for rings with large length.
%The formula in eq.~(\ref{relation}) is derived from physics.
Below, we derive the formula in eq.~(\ref{relation}).
First, consider the local Hamiltonian of the $i$-th ring,
\begin{equation}
H^{(i)}_\rm{ring}(\{s_{i,j} \})=-J_\rm{F}(i) \sum_{j=1}^{L(i)} s_{i,j} s_{i,j+1}.
\label{ring_Hamiltonian}
\end{equation}
Here, the effect due to inter-ring couplings between rings $\{J_{ij}\}$ and the biases on spins $\{h_i/L(i)\}$ is neglected.
The correlation length $\xi_i(T)$ of this model at temperature $T$ is given by~\cite{nishimori2010elements} (see Appendix~\ref{Appendix1} for a detailed derivation)
\begin{equation}
\xi_i(T)^{-1} =- \log \left[ \tanh \left(\frac{J_{\rm F}(i)}{T} \right) \right],
\label{correlation}
\end{equation}
where $\xi_i(T)$ is defined by
\begin{equation}
C_i(j)=\langle s_{i,k} s_{i, j+k} \rangle_T = \exp \left( -\frac{j}{\xi_i(T)}\right).
\end{equation}
Here, $C_i(j)$ is called the correlation function.
It describes the thermal average of the products of spins $s_{i,k}$ and $s_{i, j+k}$.
The correlation function is independent of $k$ due to the periodic boundary condition of the ring.
As the distance between the spins increases, the value of $C_i(j)$ decays exponentially.
The correlation length $\xi_i(T)$ determines the decay length scale.
$\xi_i(T)$ is a monotonically decreasing function of $T$.
At sufficiently low temperatures, the correlation length is much larger than the ring length $\xi_i(T) \gg L(i)$.
Hence, all the spins in the ring tend to have the same values.
On the other hand, at sufficiently high temperatures, $\xi_i(T) \ll L(i)$.
In this case, each spin in the ring randomly has values $+1$ or $-1$.
The crossover occurs at $T_{\rm c}(i)$, where
\begin{equation}
L(i)=\xi_i(T_\rm{c}(i)).
\label{T_c}
\end{equation}
Here, we assume that $T_{\rm c}(i)$ of all the rings have the same value.
That is,
\begin{equation}
    T_\rm{c}(i)=J_\rm{c}.
    \label{principle}
\end{equation}
Substituting eqs.~(\ref{correlation}) and (\ref{principle}) into (\ref{T_c}) gives (\ref{relation}).
\end{itemize}

The guiding principle of ME design to achieve a high performance in Ising machines is that the intra-ring-coupling strength must be tuned according to eq.~(\ref{relation}).
In SA, the temperature $T$ decreases from a high temperature to a low temperature.
The physical spins in each ring randomly take values of $+1$ or $-1$ when $T > J_\rm{c}$.
By contrast, they are aligned in the same direction when $T < J_\rm{c}$.
The physical spins in each ring are aligned along the same direction simultaneously at $T=J_\rm{c}$.

When the lengths of rings are uniform and $L(i)=N_{\rm L}-1$, ME~$iii$ is reduced to ME~$i$.
As such, the case with uniform-length and nonuniform-coupling ME is not considered in this study.
Next, we compared the performance of the three MEs.

\section{Experimental setup}\label{Sec:Experiment}
\subsection{Benchmarking problems} \label{Subsec:benchmarking}
We considered four types of benchmarking problems (i.e., logical Ising models). 
Each benchmarking problem has its own distribution of the degree in $G_{\rm L}$ or of $\{ J_{ij} \}$ and $\{ h_i \}$.

\begin{itemize}
\item{\it Binomial-Bimodal problem}\\
The logical graph $G_{\rm L}$ is created by connecting the vertices $i$ and $j$ by an edge with half probability.
The degree distribution is given by the binomial distribution.
The coupling strengths and the biases are chosen according to a bimodal distribution.
That is, $J_{ij}$ and $h_i$ take values from $\{-1, +1\}$ with equal probability.

\item{\it Binomial-Gaussian problem}\\
The logical graph $G_{\rm L}$ is created by connecting the vertices $i$ and $j$ by an edge with half probability.
The coupling strengths and the biases are chosen according to a Gaussian distribution with a mean of zero and a standard deviation of unity.

\item{\it Power-Bimodal problem}\\
The logical graph $G_{\rm L}$ with a scale-free network is created by the algorithm of the Barabasi-Albert (BA) model~\cite{albert2002statistical}.
The degree distribution is given by a power-law distribution.
The coupling strengths and the biases are chosen according to a bimodal distribution.
That is, $J_{ij}$ and $h_i$ take values from $\{-1, +1\}$ with an equal probability.

\item{\it Power-Gaussian problem}\\
The logical graph $G_{\rm L}$ with a scale-free network is created by the algorithm of the BA model.
The coupling strengths and the biases are chosen according to a Gaussian distribution with a mean of zero and a standard deviation of unity.\\
\end{itemize}

\begin{figure}[!t]
\centering
\includegraphics[width=\linewidth]{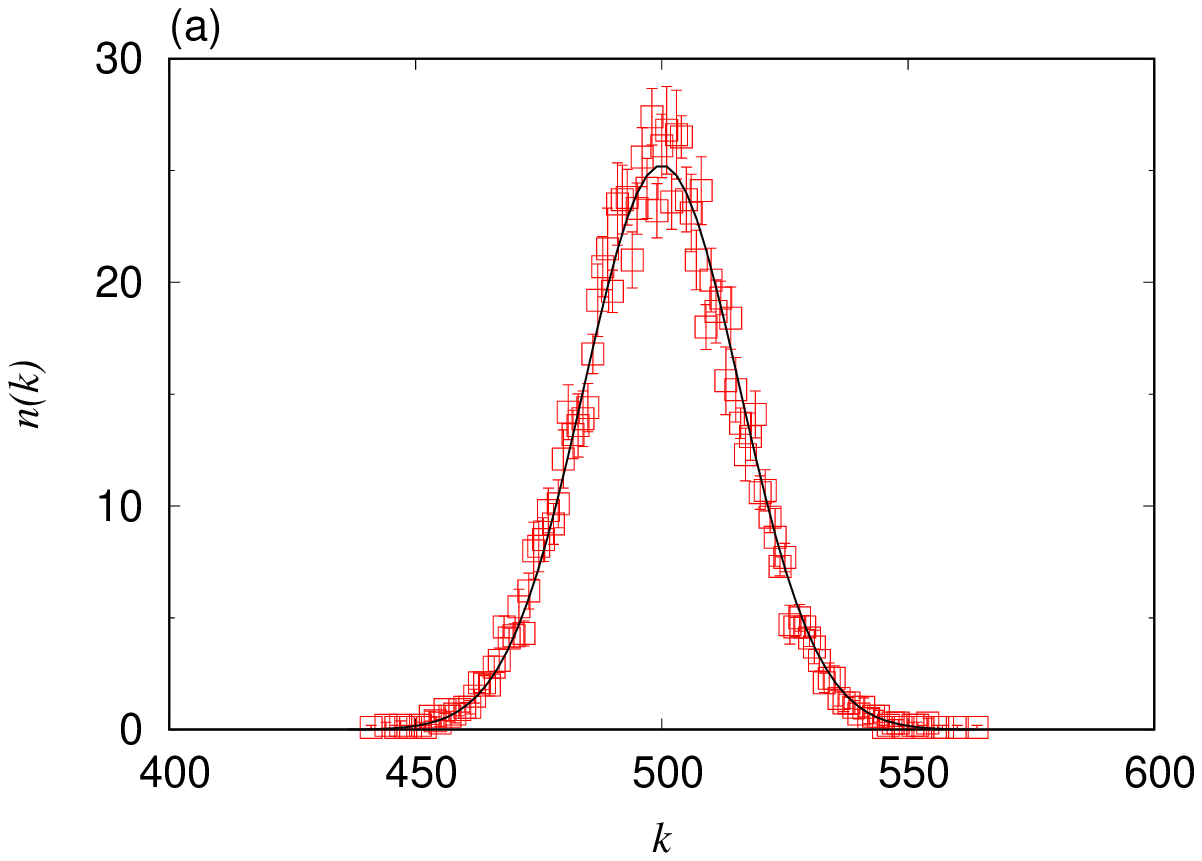}
\includegraphics[width=\linewidth]{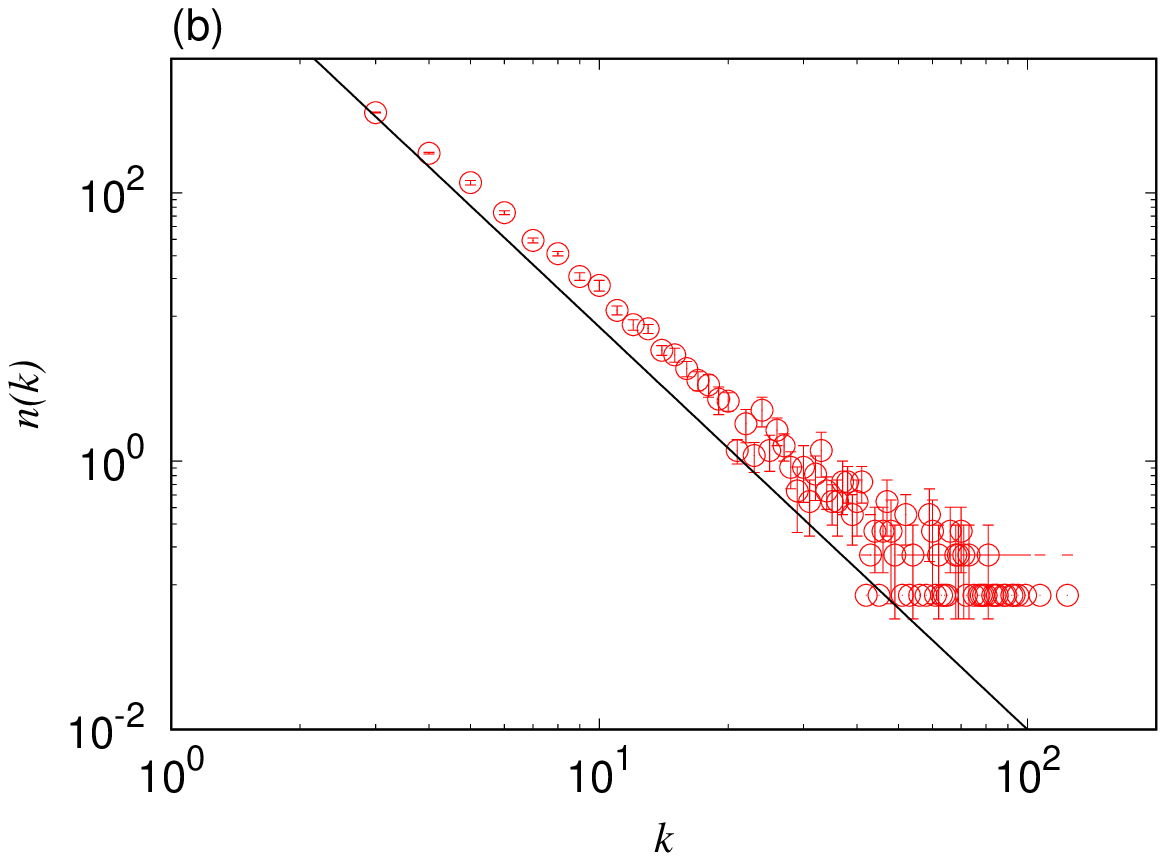}
\caption{Histograms of the degree for the logical Ising model with $1000$ spins in (a) the Binomial-Bimodal problem and the Binomial-Gaussian problem and (b) the Power-Bimodal problem and the Power-Gaussian problem.
In Fig.~(a), the line is drawn using a Gaussian distribution.
In Fig.~(b), the line is a guide to show the power-law scaling of $n(k) \propto k^{-3}$.
Error bars are the standard deviation of $10$ realizations of the logical graphs for each benchmarking problem.
}
\label{fig_degree}
\end{figure}

For each benchmarking problem, we prepared a hundred scenarios by creating ten connected logical graphs.
For each connected graph, we generated ten sets of different coupling strengths $\{ J_{ij} \}$ and biases $\{ h_i\}$.

The BA model was originally introduced to explain the mechanism responsible for the emergence of power-law degree distributions of networks in various fields.
In the algorithm of the BA model, the graph begins from a fully connected graph with $m_0$ vertices.
At every step, a new vertex with $m$ edges is added to $m$ different vertices already present in the graph with a certain probability.
The probability of connecting a new vertex and vertex $i$ depends on the degree $k_i$, and is given as
\begin{equation}
P(i)=\frac{k_i}{\sum_j k_j}.
\end{equation}
The algorithm ends when the number of vertices is $N_{\rm L}$.
Since the number of edges increases by $m$ in every step, the number of edges in a graph with $N_{\rm L}$ vertices is approximately
\begin{equation}
    \sum_{i\in V_\rm{L}} k_i \simeq m N_{\rm L}.
    \label{edge_Power}
\end{equation}
Numerical simulations and analytic results~\cite{albert2002statistical} have demonstrated that the graph evolves into a scale-free network.
Namely, the histogram of the degree $k$ denoted by $n(k)$ follows a power-law scaling.
In the BA model, the exponent is $3$ and is independent of $m_0$ and $m$.

Figure~\ref{fig_degree} shows histograms of the degree $k$ in the Binomial-Bimodal problem and the Binomial-Gaussian problem [Fig.~\ref{fig_degree} (a)] and in the Power-Bimodal problem and the Power-Gaussian problem [Fig.~\ref{fig_degree} (b)] for the model with $N_{\rm L}=1000$.
The error bars denote the standard deviation of the $10$ realizations of the logical graphs in each benchmarking problem.
In the Binomial-Bimodal problem and the Binomial-Gaussian problem, there is a peak around $k=N_{\rm L}/2$ because the vertices are connected by an edge with half probability.
The peak width is the order of $N_{\rm L}^{1/2}$.
The histogram is well described by a scaled Gaussian distribution with a mean of $N_\rm{L}/2$ and a standard deviation of $\sqrt{N_\rm{L}}/2$.
On the other hand, in the Power-Bimodal problem and the Power-Gaussian problem, the degree is more widely distributed, and the histogram follows a power law.
The power-law scaling of $n(k) \propto k^{-3}$ is consistent with our data.
Here, we set $m_0=3$ and $m=3$.

\subsection{Simulation details}\label{Subsec:simulation}
We applied SA to the physical Ising models by adopting the single-spin flip Monte Carlo method.
In each update of the spin configuration, the spin is randomly selected and the energy difference $\Delta E$ is calculated between the current state and the candidate state in which the chosen spin is flipped [see eq.~(\ref{energy_diff})].
Here, the heat-bath transition probability at temperature $T$ is used and is given as
\begin{equation}
W(\Delta E, T)=\left[ 1+\exp \left(\frac{\Delta E}{T} \right) \right]^{-1}.
\label{heatbath}
\end{equation}
Equation~(\ref{heatbath}) satisfies the balance condition [see eq.~(\ref{detailed})], and each Monte Carlo step (MCS) repeats the updates $|V_{\rm P}|$ times.
The temperature is initially set to $T_\rm{ini}=10$, which is larger than the typical energy scale, and decreases by $10^{-4}$ in every MCS.
The temperature at the end of the annealing is zero.
Appendix~\ref{Appendix2} shows the result using a different type of annealing schedule.
Regardless of the annealing schedule, the same results are qualitatively produced.

After performing SA, the values of the logical spins $\{ \sigma_i\}$ are determined from the spin configuration of the physical Ising model. 
If all the physical spins in the ring have the same value, the value is the same as that for the logical spin.
If not, the value of the logical spin $+1$ or $-1$ is determined by the majority vote.
Namely, when five physical spins take $+1$ and three physical spins take $-1$ in a ring $\phi(i)$, $\sigma_i$ is determined as $+1$. 
If $(+1)$-spins and $(-1)$-spins are the same, the value of the corresponding logical spin is set to $+1$.

For each physical Ising model (i.e., the model mapped by an embedding), we performed SA one hundred times to estimate the average and standard deviation of the quantities described in Sec.~\ref{Numerical}.

%\subsection{Indicators for the performance of the embedding algorithm}~\label{Subsec:indicators}

\section{Numerical Results}~\label{Numerical}
\begin{figure}[!t]
\centering
\includegraphics[width=\linewidth]{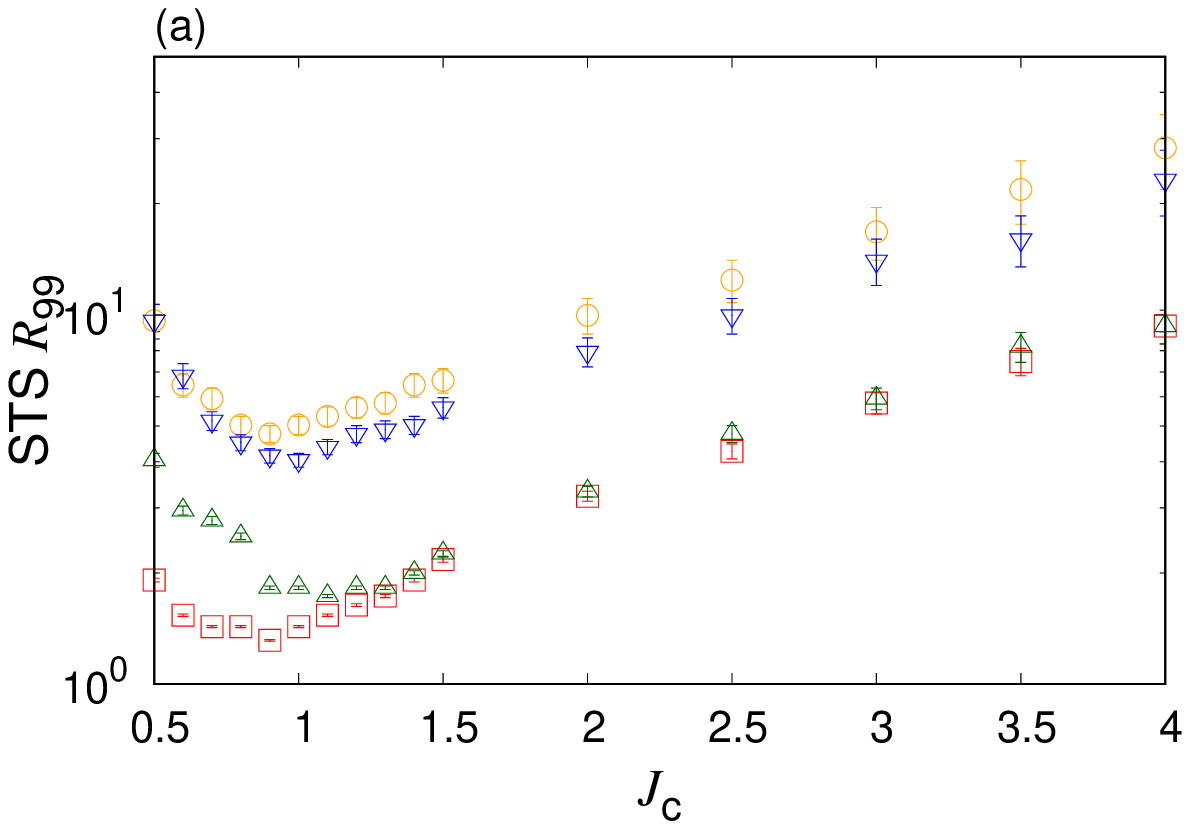}\\
\includegraphics[width=\linewidth]{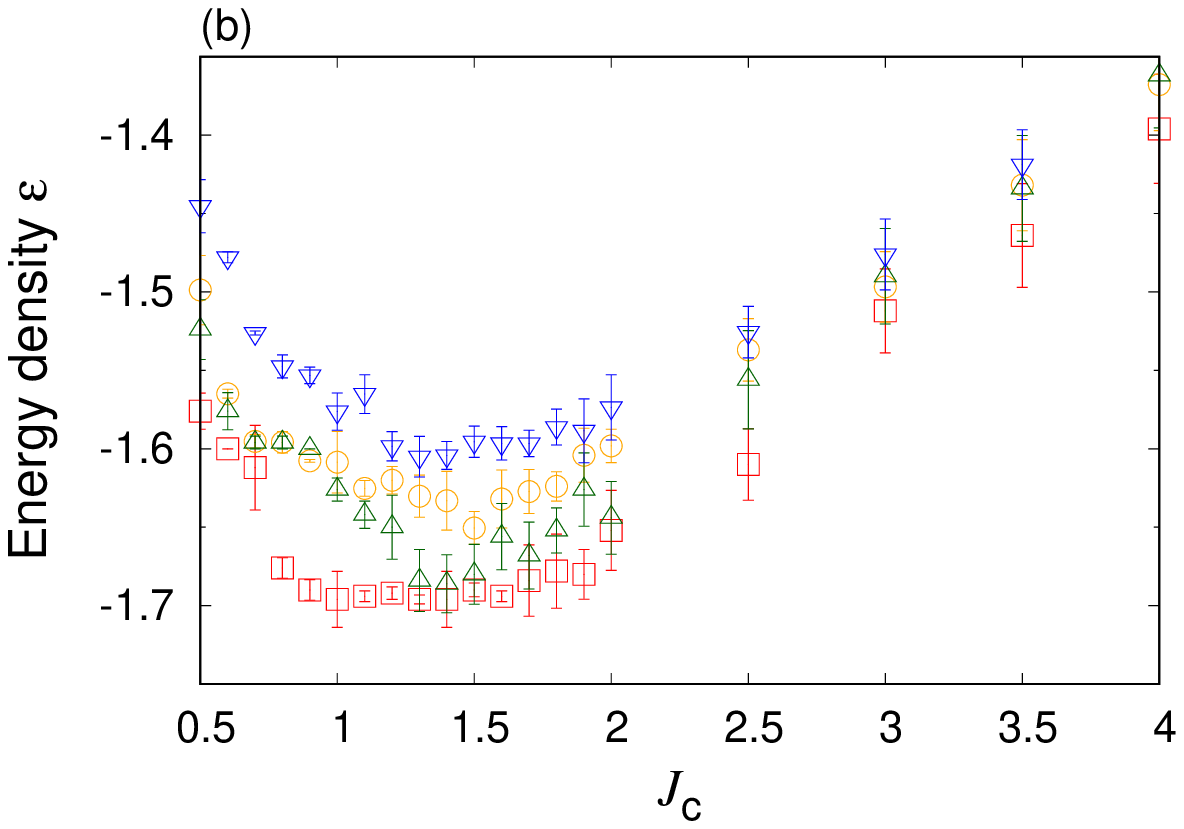}
\caption{
(a) $J_\rm{c}$-dependences of the STS $R_{99}$ and (b) the energy density $\epsilon$ in ME~$i$.
Squares (red), circles (orange), up-triangles (green), and down-triangles (blue) denote the Binomial-Bimodal problem, the Binomial-Gaussian problem, the Power-Bimodal problem, and the Power-Gaussian problem, respectively.
Number of logical spins used is (a) $N_{\rm L}=8$ and (b) $N_{\rm L}=10$.
}
\label{Fig:JF_dep}
\end{figure}

We compared the performances of ME~$i$, ME~$ii$, and ME~$iii$ for each benchmarking problem.
We measured two quantities.
The first one is the step to solution (STS).
The STS is the number of steps required for the algorithm to obtain the ground state at least once with a probability of $0.99$, and it is defined by~\cite{boixo2014evidence, aramon2019physics}
\begin{equation}
R_{99}^{(k)}=\frac{\log(1-0.99)}{\log(1-{\rm P}^{(k)}_{\rm s})},
\label{STS}
\end{equation}
where ${\rm P}_{\rm s}^{(k)}$ is the success probability and $k$ is a label of the logical Ising models.
Herein $k$ is the run, which ranges from $1$ to $100$.
A small $R_{99}^{(k)}$ value indicates a good performance.
Here, the success probability is estimated as ${\rm P}^{(k)}_{\rm s}= N_{\rm s}^{(k)}/100$, where $N_{\rm s}^{(k)}$ is the number of obtained ground states in one hundred simulations of SA.
We measured STS for relatively small-size systems up to $N_{\rm L}=28$ because it is difficult to obtain the ground state of $H_{\rm L}(\{\sigma_i\})$ for a larger system size.

For the $N_{\rm L}$-dependence of the performance in a larger-sized system, we calculated the energy density (i.e., the value of $H_{\rm L}(\{\sigma_i\})/N_{\rm L}$), where $\sigma_i$ is determined by the majority vote after SA (see subsection~\ref{Subsec:simulation}).
$\epsilon^{(k)}$ denotes the average of the energy density for 100 simulations of SA, where $k$ is the label of the logical Ising model.
A smaller $\epsilon^{(k)}$ indicates a better performance.
Note that the energy density can be evaluated on the order of $N_{\rm L}^2$ steps.
Thus, this quantity is useful to study larger-sized systems.

To investigate the performance of ME in the benchmarking problems, we used the median of the STSs and the median of energy densities.
These are denoted as $R_{99}$ and $\epsilon$, respectively.

Figure~\ref{Fig:JF_dep} plots the $J_\rm{c}$-dependences of $R_{99}$ (a) and $\epsilon$ (b) in ME~$i$ for each benchmarking problem.
Here, the optimal values of $J_\rm{c}$ that minimizes $R_{99}$ or $\epsilon$ are found.
We estimated the optimal values of $J_\rm{c}$ in each ME for different sized systems, where 0.1 is used as the precision threshold of $J_\rm{c}$ (see Appendix~\ref{Appendix3} for the $N_{\rm L}$-dependences).
In the following, we show the data of $R_{99}$, $\epsilon$, and $\epsilon^{(k)}$ at the optimal value of $J_\rm{c}$.

\subsection{Step to Solution (STS)}
\begin{figure*}[!t]
\centering
\includegraphics[width=\linewidth]{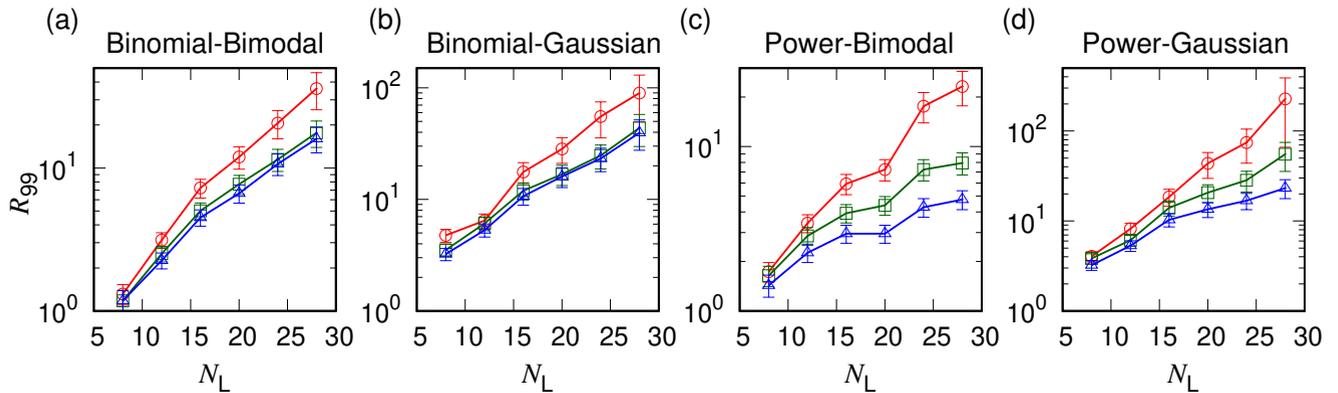}
\caption{
$N_{\rm L}$-dependences of the STS for (a) the Binomial-Bimodal problem, (b) the Binomial-Gaussian problem, (c) the Power-Bimodal problem, and (d) the Power-Gaussian problem.
ME~$i$, ME~$ii$, and ME~$iii$ are denoted by red circles, green squares, and blue triangles, respectively.
}
\label{fig_STS}
\end{figure*}

First, we used STS to compare the performance among ME~$i$, ME~$ii$, and ME~$iii$.
Figure~\ref{fig_STS} shows the $N_\rm{L}$-dependences of STS for each benchmarking problem.
In all cases, STS increases exponentially with $N_\rm{L}$.
For small system sizes, $N_{\rm L}=8$ or $12$, there is not a clear difference in performance.
However, a clear difference appears as the number of logical spins increases.
For all benchmarking problems, ME~$i$ shows the poorest performance.
For the Binomial-Bimodal problem [Fig.~\ref{fig_STS} (a)] and the Binomial-Gaussian problem [Fig.~\ref{fig_STS} (b)], the performances of ME~$ii$ and ME~$iii$ are the same within the margin of error.
On the other hand, for the Power-Bimodal problem [Fig.~\ref{fig_STS} (c)] and the Power-Gaussian problem [Fig.~\ref{fig_STS} (d)], ME~$iii$ outperforms ME~$ii$.

We also considered the time to solution (TTS)~\cite{boixo2014evidence, ronnow2014defining, ohzeki2019control, aramon2019physics}.
TTS is the {\it total time} required for the algorithm to obtain the ground state at least once with a probability of $0.99$.
TTS is estimated as
\begin{equation}
    \rm{TTS} \simeq n_\rm{MCS} \times \tau_\rm{MCS} \times R_{99},
\end{equation}
where $n_\rm{MCS}$ is the number of the MCSs in the SA and $\tau_\rm{MCS}$ is the time required to calculate one MCS.
%In the following, we give an argument that the TTS shows qualitatively the same result as the STS.
For all three MEs, $n_\rm{MCS}$ is the same as STS and is given as
\begin{equation}
    n_\rm{MCS}=10^5.
\end{equation}
In each MCS, we calculated the energy difference of a single-spin flip $|V_{\rm P}|$ times.
The time required for the calculation of the energy difference is $O(|V_{\rm P}|^0)$ since the connectivity of the physical Ising model is sparse.
Thus
\begin{equation}
    \tau_\rm{MCS}\sim |V_\rm{P}|.
\end{equation}
In the Binomial-Bimodal problem and the Binomial-Gaussian problem, the number of physical spins is on the order of $N_{\rm L}^2$ for all three MEs.
Thus, TTS qualitatively shows the same result as STS.
In the Power-Bimodal problem and the Power-Gaussian problem, the number of physical spins is on the order of $N_{\rm L}^2$ in ME~$i$, while it is on the order of $N_{\rm L}$ in ME~$ii$ and ME~$iii$.
The scaling of $O(N_{\rm L})$ is obtained from eqs.~(\ref{vertex_Power}) and (\ref{edge_Power}).
TTS shows larger performance differences between ME~$i$ and the other two compared to the STS.

\subsection{Energy density}
\begin{figure*}[!t]
\centering
\includegraphics[width=\linewidth]{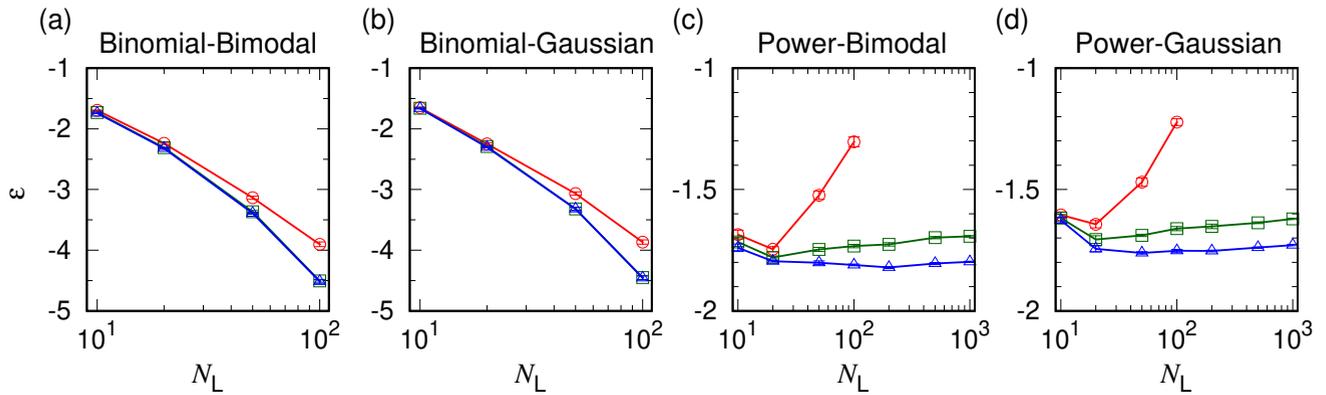}
\caption{
$N_{\rm L}$-dependences of the energy density for (a) the Binomial-Bimodal problem, (b) the Binomial-Gaussian problem, (c) the Power-Bimodal problem, and (d) the Power-Gaussian problem.
ME~$i$, ME~$ii$, and ME~$iii$ are denoted by red circles, green squares, and blue triangles, respectively.
In Figs.~(a) and (b), the data of ME~$ii$ and ME~$iii$ overlap with each other.
}
\label{fig_Energy}
\end{figure*}

\begin{figure*}[!t]
\centering
\includegraphics[width=\linewidth]{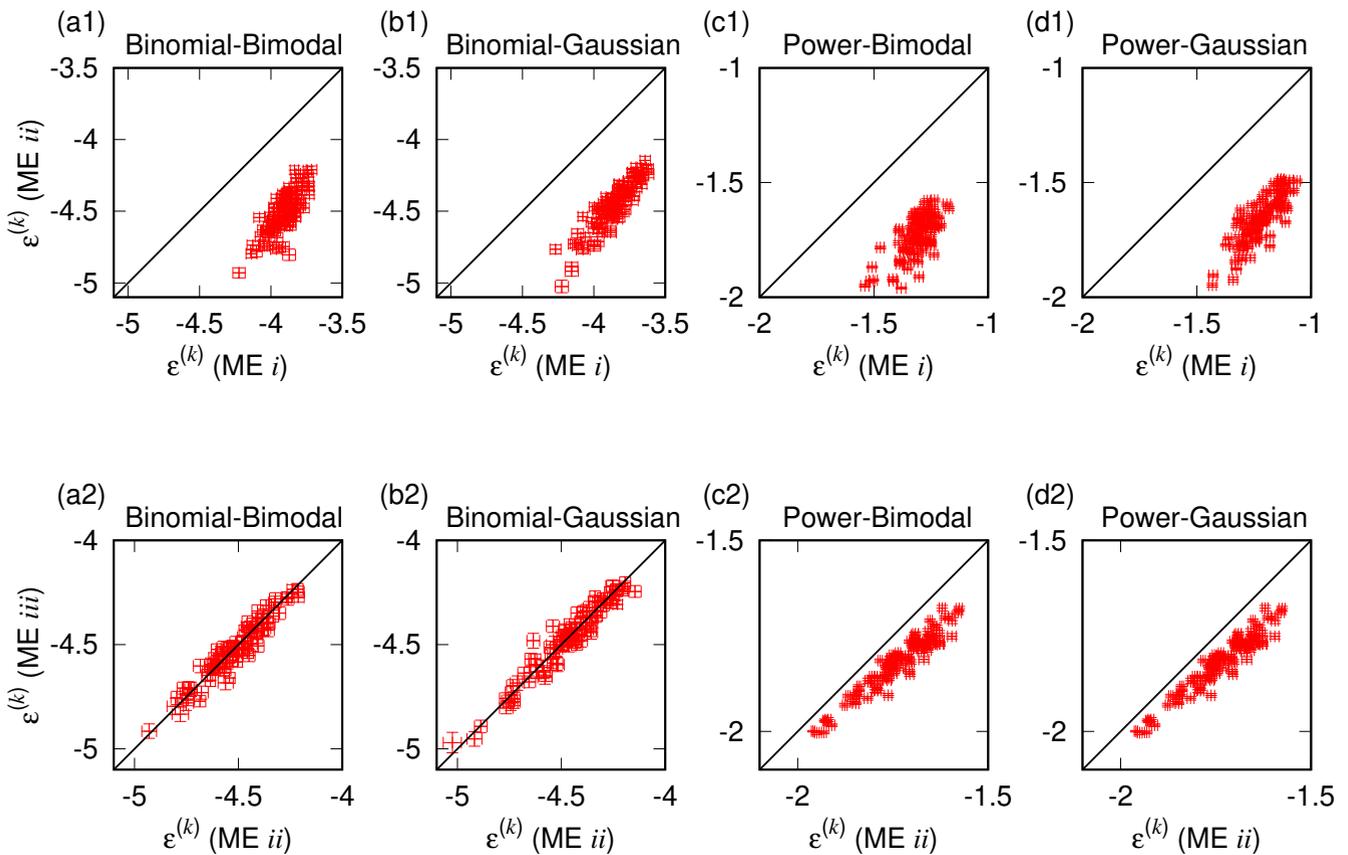}
\caption{
Scatterplots of the energy densities $\{ \epsilon_k\}_{k=1}^{100}$ at $N_{\rm L}=100$ among ME~$i$, ME~$ii$, and ME~$iii$ for (a) the Binomial-Bimodal problem, (b) the Binomial-Gaussian problem, (c) the Power-Bimodal problem, and (d) the Power-Gaussian problem.
Upper panel compares ME~$i$ and ME~$ii$, and the lower panel compares ME~$ii$ and ME~$iii$.
}
\label{fig_Evs}
\end{figure*}

Next, we compared the performance among ME~$i$, ME~$ii$, and ME~$iii$ in terms of the energy density.
The results are qualitatively the same as those of STS.
Figure~\ref{fig_Energy} shows the $N_{\rm L}$-dependence of $\epsilon$ for each benchmarking problem.
As $N_{\rm L}$ increases, the difference in performance among the MEs becomes clear.
ME~$i$ has the poorest performance.
ME~$ii$ and ME~$iii$ have the same performance for the Binomial-Bimodal problem [Fig.~\ref{fig_Energy} (a)] and the Binomial-Gaussian problem [Fig.~\ref{fig_Energy} (b)].
On the other hand, ME~$iii$ outperforms ME~$ii$ for the Power-Bimodal problem [Fig.~\ref{fig_Energy} (c)] and the Power-Gaussian problem [Fig.~\ref{fig_Energy} (d)].
Figures~\ref{fig_Energy} (c) and (d) show that the differences in energy densities are almost the same for $N_{\rm L} \geq 10^2$.
This implies that ME~$iii$ will provide the best performance, even for larger-sized systems.

Figure~\ref{fig_Evs} shows the scatterplot to compare the energy densities for each benchmarking problem (i.e., $\{\epsilon_k\}_{k=1}^{100}$, among the three MEs) using the data at $N_\rm{L}=100$.
The upper panel compares the energy densities between ME~$i$ and ME~$ii$.
All the points are plotted below the diagonal, indicating that ME~$ii$ outperforms ME~$i$ for all the logical Ising models in each benchmarking problem.
The lower panel compares the energy densities between ME~$ii$ and ME~$iii$.
For the Binomial-Bimodal problem [Fig.~\ref{fig_Evs} (a2)] and the Binomial-Gaussian problem [Fig.~\ref{fig_Evs} (b2)], the points are plotted around the diagonal, implying that ME~$ii$ and ME~$iii$ have similar performance.
On the other hand, for the Power-Bimodal problem [Fig.~\ref{fig_Evs} (c2)] and the Power-Gaussian problem [Fig.~\ref{fig_Evs} (d2)], all the points are plotted below the diagonal, indicating that ME~$iii$ is better suited for these benchmarking problems.

\section{Discussion}\label{Sec:Discussion}
The numerical studies demonstrate that ME~$i$ has the poorest performance.
The poor performance of ME~$i$ is attributed to the large dimension of the solution space.
In ME~$i$, the logical Ising model with $N_{\rm L}$-spins is mapped to the physical Ising model with $N_{\rm L}(N_{\rm L}-1)$-spins.
On the other hand, in ME~$ii$, the number of the physical spins is about $N_{\rm L}^2/2$ for the Binomial-Bimodal problem and the Binomial-Gaussian problem, but is on the order of $N_{\rm L}$ for the Power-Bimodal problem and the Power-Gaussian problem.
The dimension of the solution space increases exponentially with respect to the number of physical spins.
Thus, the dimension of the solution space in ME~$i$ rapidly increases with $N_{\rm L}$ compared to those in ME~$ii$ and ME~$ii$.

ME~$ii$ and ME~$iii$ have the same performance for the Binomial-Bimodal problem and the Binomial-Gaussian problem.
This can be understood as follows for large $L(i)$.
For the Binomial problem, the degree distribution shows a peak around $k=N_{\rm L}/2$ with the width on the order of $N_{\rm L}^{1/2}$ [see Fig.~\ref{fig_degree}(a)].
In ME~$iii$, the length of the ring is equal to the degree.
Thus, $L(i)$ is distributed with a mean of $L_\rm{mean}=N_{\rm L}/2$ and a standard deviation $\Delta L$ on the order of $N_{\rm L}^{1/2}$.
For large $L(i)$, the intra-ring-coupling strength $J_{\rm F}(i)$ behaves as the $\log L(i)$ [see eq.~(\ref{Log})].
Then, the standard deviation of $J_{\rm F}(i)$ scales as
\begin{equation}
\Delta J_\rm{F} \sim \Delta (\log L) \sim  \frac{\Delta L}{L_\rm{mean}} \sim N_\rm{L}^{-\frac{1}{2}}.
\end{equation}
The standard deviation $\Delta J_{\rm F}$ decays with $N_{\rm L}$, implying that ME~$iii$ approaches ME~$ii$ as the system size increases.

On the other hand, ME~$iii$ outperforms ME~$ii$ for the Power-Bimodal problem and the Power-Gaussian problem for large $N_{\rm L}$.
In these problems, the degree of the logical Ising model is widely distributed, reflecting the distribution of $L(i)$ in the physical Ising model.
Namely, the lengths of some rings are on the order of $1$, while others are on the order of $N_{\rm L}$.
In these problems, it is necessary to set the intra-ring-coupling strength according to eq.~(\ref{relation}) to achieve a high performance in SA-based Ising machines.

\section{Conclusion and Outlook}\label{Sec:conclusion}
Here, we discussed the guiding principle of ME design to achieve a high performance in SA-based Ising machines from a viewpoint of statistical mechanics.
We proposed a new type of ME shown in eq.~(\ref{relation}).
In the proposed ME, the coupling strength inside a chain depends on the chain length.
This is a unique approach that has not been discussed previously.
We compared the performance of our proposed ME with the two existing MEs using four benchmarking problems.
SA showed that the proposed ME has the best performance for all the benchmarking problems.
In particular, it outperformed the others when the logical Ising model has a wide degree distribution.
The results are independent of the distribution of coupling strengths and biases in the logical Ising model.

We demonstrated the importance of tuning the intra-chain coupling strengths in SA, which is regarded as an ideal Ising machine.
In the future, we plan to apply eq.~(\ref{relation}) to real Ising machines such as a CMOS annealing machine.

It is also important to compare our results with the case of quantum annealing (QA)~\cite{kadowaki1998quantum}, where the transverse-field strength $\Gamma$ plays the role of the temperature $T$ in SA.
A recent paper~\cite{hamerly2019experimental} shows that ME~$ii$ outperforms ME~$i$.
This is consistent with the results in this study.
Furthermore, our results imply that the performance of QA could be enhanced for some problems if the intra-chain coupling strength $J_{\rm F}$ is tuned according to the chain length.
The D-Wave's report~\cite{andriyash2016boosting} evaluated the chain-length dependence of the tunneling energy between the all-up-spin state and the all-down-spin state of chains, and showed that $\Gamma/J_{\rm F}$ should be larger for a longer chain to achieve a high performance of QA.
In this study, $\Gamma$ was tuned instead of $J_{\rm F}$.
Interestingly, their result implied the opposite as ours using SA.
This is a future problem to uncover the origin of the difference between QA and SA.

\appendices
\section{Correlation length in a one-dimensional Ising model}\label{Appendix1}
The appendix provides a detailed derivation of the correlation length in a one-dimensional Ising model [see eq.~(\ref{correlation}) in the main text].
The correlation length is determined by the correlation function, which is given by
\begin{align}
C_i (j)=&\langle s_{i,1} s_{i,j+1} \rangle_T, \nonumber \\
=& \sum_{\{s_{i,l}\}} s_{i,1} s_{i,j+1} {\rm P}_\rm{eq}(\{s_{i,l}, T \}), \nonumber\\
=&\sum_{\{s_{i,l}\}} s_{i,1} s_{i,j+1} \exp{\left(K(i)\sum_{k=1}^{L(i)} s_{i,k} s_{i,k+1}\right)}/Z.
\label{correlation_appendix}
\end{align}
We use the Hamiltonian shown in eq.~(\ref{ring_Hamiltonian}) and set $K(i)=J_{\rm F}(i)/T$.
Here, $Z$ is called the partition function in statistical mechanics, which is given by
\begin{equation}
    Z=\sum_{\{s_{i,l}\}} \exp{\left(K(i)\sum_{k=1}^{L(i)} s_{i,k} s_{i,k+1}\right)}.
\end{equation}

The transfer matrix method is a powerful tool in statistical mechanics.
We applied it to calculate the partition function $Z$ and the numerator on the right-hand side of eq.~(\ref{correlation_appendix}).
First, we introduce
\begin{equation}
    T(s_{i,j}, s_{i,j+1})=\exp{(K(i)s_{i,j}s_{i,j+1})}.
\end{equation}
Then the partition function is given by
\begin{align}
    Z=\sum_{l=1}^{L(i)} \sum_{s_{i,l}\in\{+1,-1\}} T(s_{i,1},s_{i,2}) T(s_{i,2},s_{i,3}) \cdots \nonumber\\ \cdots T(s_{i,L(i)-1},s_{i,L(i)}) T(s_{i,L(i)},s_{i,1}).
\end{align}
Here, it is convenient to regard $T(s_{i,j}, s_{i,j+1})$ as a matrix element of ${\rm T}$ such as
\begin{equation}
    {\rm T}=
    \begin{pmatrix}
    T(1,1)&T(1,-1)\\
    T(-1,1)&T(-1,-1)
    \end{pmatrix}
    =
    \begin{pmatrix}
    e^{K(i)}&e^{-K(i)}\\
    e^{-K(i)}&e^{K(i)}
    \end{pmatrix}
    .
\end{equation}
The matrix ${\rm T}$ is called the transfer matrix.
Then the partition function is written as
\begin{equation}
    Z= \rm{Tr} \left(\rm{T}^{L(i)}\right)=\lambda_+^{L(i)} + \lambda_-^{L(i)},
\end{equation}
where $\lambda_\pm$ are the two eigenvalues of ${\rm T}$ [i.e., $\lambda_+=2\cosh K(i)$ and $\lambda_-=2\sinh K(i)$].
Similarly, the numerator on the right-hand side of eq.~(\ref{correlation_appendix}) is evaluated as
\begin{align}
    &\sum_{\{s_{i,l}\}} s_{i,1} s_{i,j+1} \exp{\left(K(i)\sum_{k=1}^{L(i)} s_{i,k} s_{i,k+1}\right)}\nonumber\\
    =&\sum_{l=1}^{L(i)} \sum_{s_{i,l}\in\{+1,-1\}} s_{i,1} s_{i,j+1} \nonumber\\
    &\times T(s_{i,1},s_{i,2}) T(s_{i,2},s_{i,3}) \cdots T(s_{i,L(i)},s_{i,1}), \nonumber\\
    =& {\rm T}{\rm r} \left( \sigma \rm{T}^j \sigma \rm{T}^{L(i)-j} \right), \nonumber\\
    =& \lambda_+^{j} \lambda_-^{L(i)-j}+\lambda_+^{L(i)-j} \lambda_-^{j},
\end{align}
where
\begin{equation}
    \sigma=
    \begin{pmatrix}
    1&0\\
    0&-1
    \end{pmatrix}
    .
\end{equation}

The correlation function is then given by
\begin{align}
    C_i (j)=&\frac{\lambda_+^{j} \lambda_-^{L(i)-j}+\lambda_+^{L(i)-j} \lambda_-^j}{\lambda_+^{L(i)} + \lambda_-^{L(i)}},\nonumber\\
    =&\tanh^{j} K(i) \left(\frac{1+\tanh^{L(i)-2j} K(i)}{1+\tanh^{L(i)} K(i)} \right).
    \label{correlation_L}
\end{align}
For large $L(i)$, the parenthesis in eq.~(\ref{correlation_L}) can be approximated by one.
Then
\begin{equation}
    C_i(j) = \tanh^j K(i).
\end{equation}
By combining this expression with
\begin{equation}
    C_i(j)= \exp{\left(-\frac{j}{\xi_i(T)}\right)},
\end{equation}
we obtain eq.~(\ref{correlation}) in the main text.

\section{Performance comparison of Minor-Embedding for a different annealing schedule}~\label{Appendix2}
\begin{figure*}[ht]
\centering
\includegraphics[width=\linewidth]{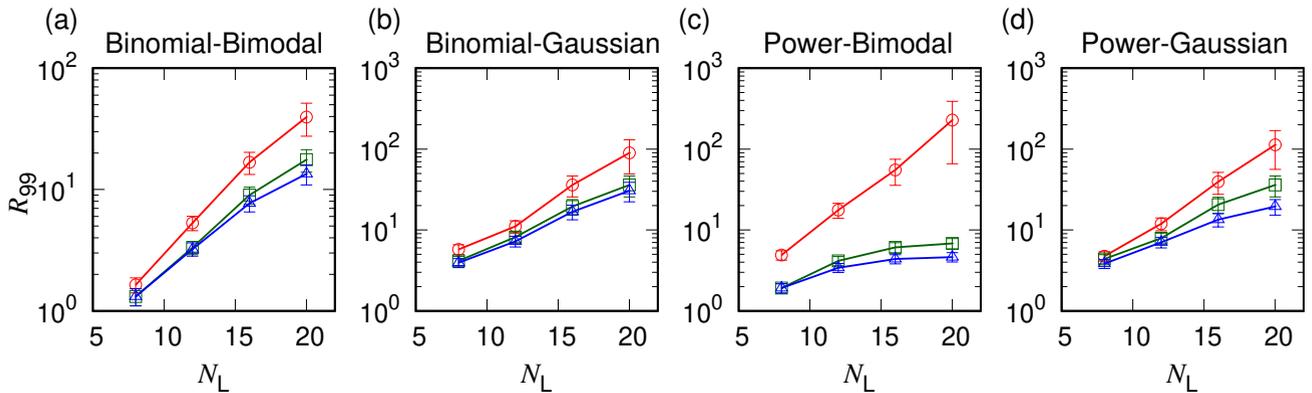}
\caption{
$N_{\rm L}$-dependences of the STS for (a) the Binomial-Bimodal problem, (b) the Binomial-Gaussian problem, (c) the Power-Bimodal problem, and (d) the Power-Gaussian problem.
ME~$i$, ME~$ii$, and ME~$iii$ are denoted by red circles, green squares, and blue triangles, respectively.
}
\label{fig_STS_v2}
\end{figure*}

\noindent
This appendix validates the proposed ME (ME~{\it iii}) when a different annealing schedule in SA is used.
Here, the temperature in the SA algorithm [see Algorithm~\ref{SA}] is taken as
\begin{equation}
    T_{k+1}=(1-r) T_{k},
    \label{schedule}
\end{equation}
where $T_k$ is the temperature at the $k$-th MCS.
We set the total number of the MCSs as $n_\rm{MCS}=10^4$, the initial temperature $T_\rm{ini}=10$, and the final temperature $T_\rm{fin}=0.01$.
The initial temperature and the final temperature are the default values in the CMOS annealing machine.
The decay rate of the temperature is determined by $T_1=T_\rm{ini}$ and $T_{n_\rm{MCS}}=T_\rm{fin}$ as $r\simeq 6.9\times 10^{-4}$.
In this annealing schedule, the temperature is lowered as an exponential function of MCSs.
This is different from the annealing schedule in the main text, where the temperature linearly decays to zero.

We use the STS given by eq.~(\ref{STS}) to compare the performances among ME~{\it i}, ME~{\it ii}, and ME~{\it iii}.
Figure~\ref{fig_STS_v2} shows the $N_\rm{L}$-dependences of the STS for each benchmarking problem.
Similar to the main text, ME~{\it iii} outperforms ME~{\it i} and ME~{\it ii} for all the benchmarking problems, implying that the results are independent of the annealing schedules.

\section{$N_{\rm L}$-dependences of the optimal values of $J_\rm{c}$}~\label{Appendix3}
\begin{figure*}[ht]
\centering
\includegraphics[width=\linewidth]{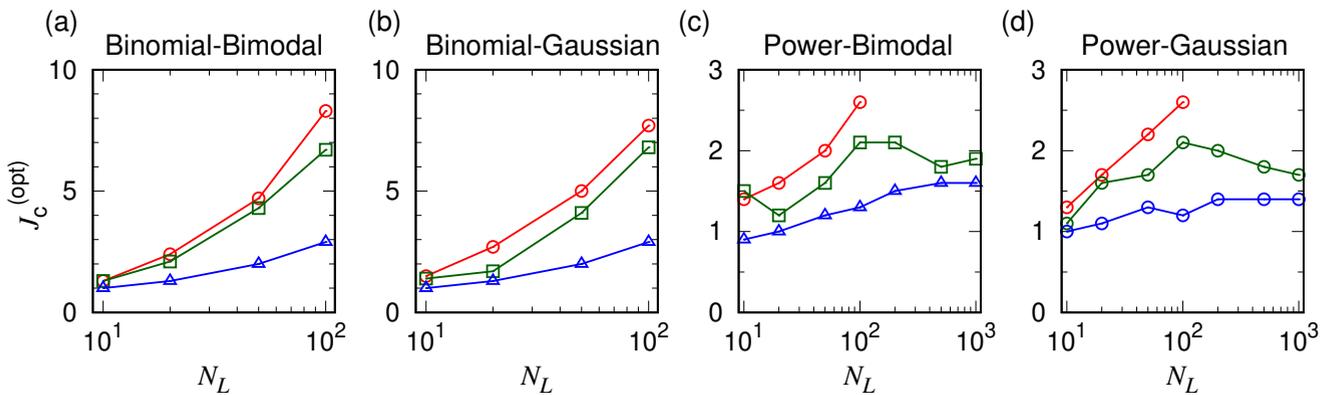}
\caption{
$N_{\rm L}$-dependences of the optimal values of $J_\rm{c}$ for (a) the Binomial-Bimodal problem, (b) the Binomial-Gaussian problem, (c) the Power-Bimodal problem, and (d) the Power-Gaussian problem.
ME~$i$, ME~$ii$ and ME~$iii$ are denoted by red circles, green squares, and blue triangles, respectively.
}
\label{fig_Opt}
\end{figure*}

\noindent
This appendix discusses the $N_\rm{L}$-dependences of the optimal value $J_\rm{c}$, which minimizes the energy density of the logical Ising model $\epsilon$.
Here, we denote the optimal value as $J_\rm{c}^\rm{(opt)}$.

Figure~\ref{fig_Opt} shows the $N_\rm{L}$-dependences of $J_\rm{c}^\rm{(opt)}$ for each benchmarking problem.
In the Binomial-Bimodal problem and the Binomial-Gaussian problem [Figs.~\ref{fig_Opt} (a) and (b)], $J_\rm{c}^\rm{(opt)}$ increases with $N_\rm{L}$.
The increase of $J_\rm{c}^\rm{(opt)}$ in ME~$iii$ is slow compared to those in ME~$i$ and ME~$ii$.
On the other hand, in the Power-Bimodal problem and the Power-Gaussian problem [Figs.~\ref{fig_Opt} (c) and (d)], the $N_\rm{L}$-dependences of the optimal values are small compared to the Binomial-Bimodal problem and the Binomial-Gaussian problem.
In ME~$i$, $J_\rm{c}^\rm{(opt)}$ increases with $N_{\rm L}$.
In ME~$ii$, $J_\rm{c}^\rm{(opt)}$ fluctuates between $J_\rm{c}^\rm{(opt)}=1$ and $J_\rm{c}^\rm{(opt)}=2$.
In ME~$iii$, $J_{\rm c}^\rm{(opt)}$ gradually increases with $N_\rm{L}$, but the increase is slower than that of the ME~$i$.

\section*{Acknowledgments}
This article is based on the results obtained from a project commissioned by the New Energy and Industrial Technology Development Organization (NEDO).
Shu Tanaka was supported in part by the Japan Science and Technology Agency (JST), PRESTO, Japan, under Grant JPMJPR1665, and in part by the Japan Society for the Promotion of Science (JSPS) KAKENHI under Grant 19H01553.
Tatsuhiko Shirai and Shu Tanaka would like to thank the Supercomputer Center, Institute for Solid State Physics, The University of Tokyo, and the supercomputers at the Yukawa Institute for Theoretical Physics, for the use of the facilities.

%One of the authors~(S.~T.~) was supported by JST, PRESTO (Grant Number JPMJPR1665), Japan and JSPS KAKENHI (Grant Number 19H01553).
%T. Shirai and S. Tanaka thank the Supercomputer Center, the Institute for Solid State Physics, The University of Tokyo and the supercomputers at the Yukawa Institute for Theoretical Physics, for the use of the facilities.

%\section*{References and Footnotes}
\bibliographystyle{IEEEtran}
\bibliography{embedding.bib}

\begin{IEEEbiography}[{\includegraphics[width=1in,height=1.25in,clip,keepaspectratio]{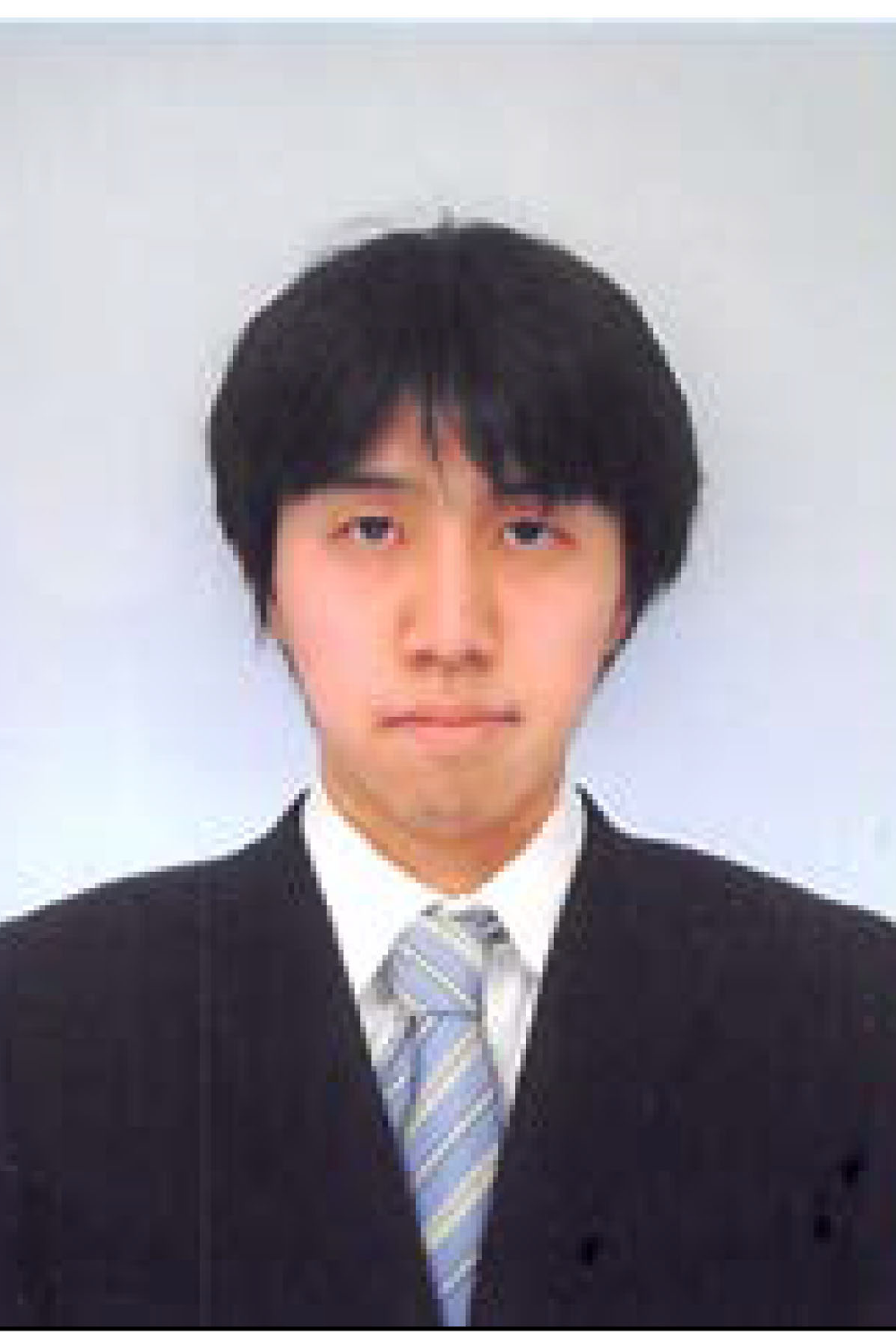}}]{Tatsuhiko Shirai}
received the B. Sci., M. Sci., and Dr. Sci. degrees from The University of Tokyo in 2011, 2013, and 2016, respectively. He is presently an assistant professor at the Department of Computer Science and Communications Engineering, Waseda University. His research interests are quantum dynamics and statistical mechanics. He is a member of JPS.
\end{IEEEbiography}

\begin{IEEEbiography}[{\includegraphics[width=1in,height=1.25in,clip,keepaspectratio]{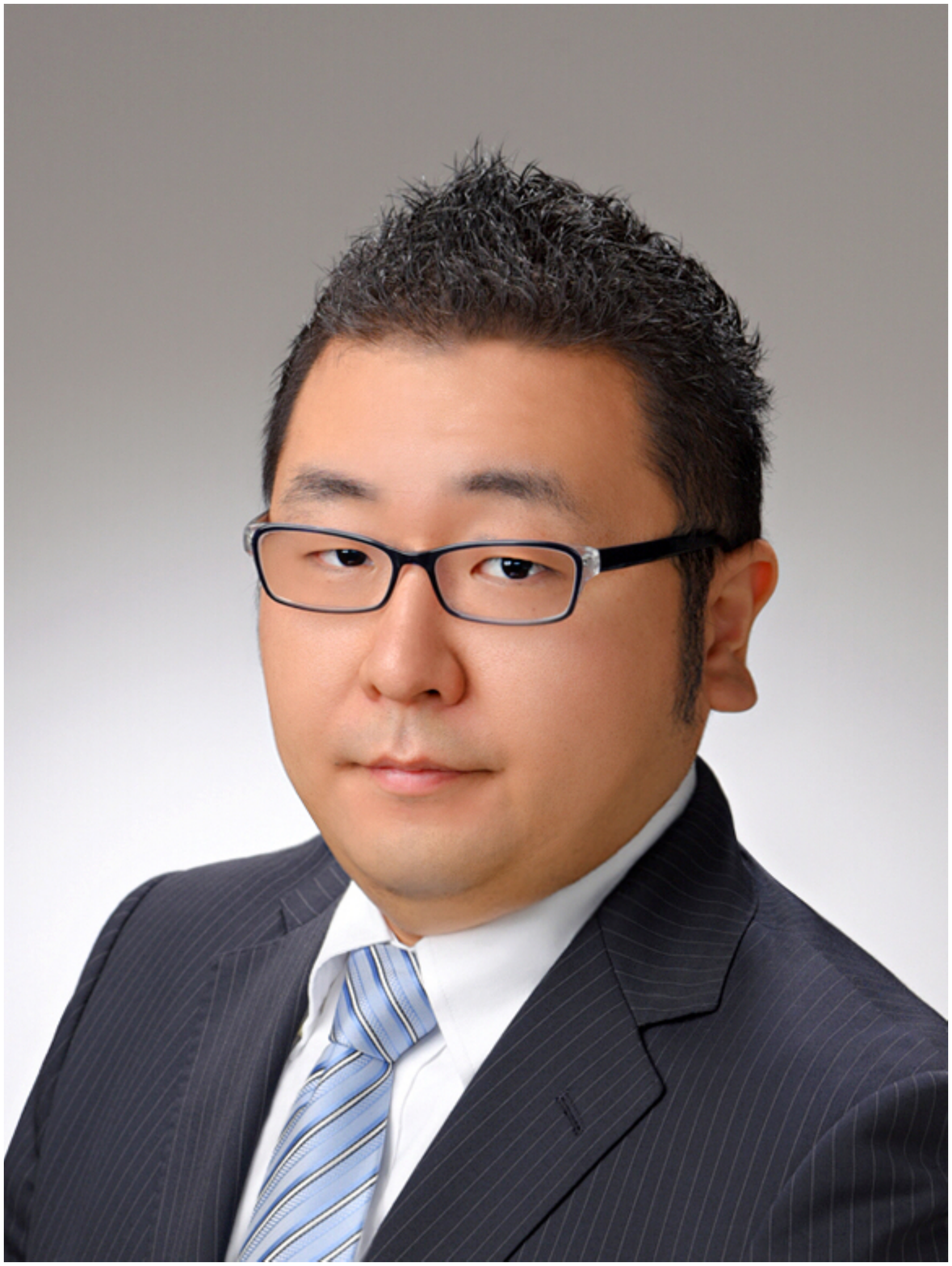}}]{Shu Tanaka}
received a B. Sci. degree from the Tokyo Institute of Technology in 2003 and the M. Sci. and Dr. Sci. degrees from The University of Tokyo in 2005 and 2008, respectively.  He is presently an associate professor in the Department of Applied Physics and Physico-Informatics, Keio University.  His research interests are quantum annealing, Ising machine, statistical mechanics, and materials science. He is a member of JPS.
\end{IEEEbiography}

%1:1.25
\begin{IEEEbiography}[{\includegraphics[width=1in,height=1.25in,clip,keepaspectratio]{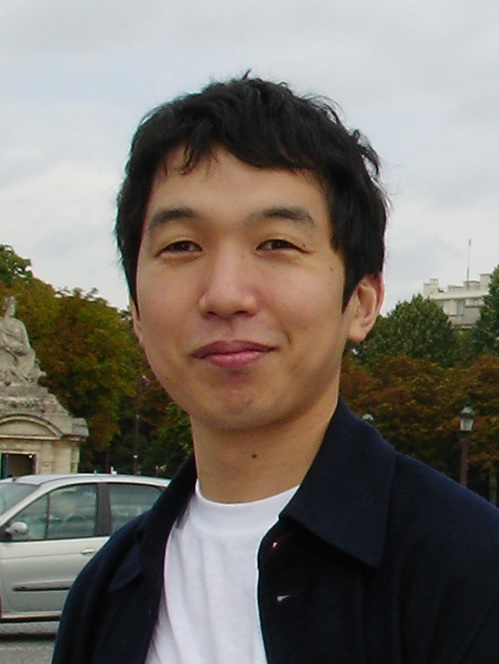}}]{Nozomu Togawa}
received  the B. Eng., M. Eng., and Dr. Eng. degrees from Waseda University in 1992, 1994,and 1997, respectively, all in electrical engineering.  He is presently a Professor in the Department of Computer Science and Communications Engineering, Waseda University. His research interests include integrated system design, graph theory, information security, and quantum computing. He is a member of IEICE and IPSJ.
\end{IEEEbiography}

\EOD

\end{document}